\documentclass[12pt, modern]{aastex62h}

\newcommand{\acronym}[1]{{\small{#1}}}
\newcommand{\Gaia}{\textsl{Gaia}}
\newcommand{\Tycho}{\textsl{Tycho}}
\newcommand{\DRone}{\textsl{\acronym{DR1}}}
\newcommand{\DRtwo}{\textsl{\acronym{DR2}}}
\newcommand{\TGAS}{\textsl{\acronym{TGAS}}}
\newcommand{\DPAC}{{\acronym{DPAC}}}
\newcommand{\documentname}{\textsl{Note}}
\newcommand{\equationname}{equation}

\newcommand{\AU}{\mathrm{A.U.}}
\newcommand{\dd}{\mathrm{d}}
\newcommand{\given}{\,|\,}
\newcommand{\T}{^{\mathsf{T}}}
\newcommand{\inv}{^{-1}}

\shorttitle{a likelihood for gaia}
\shortauthors{david w hogg}

\begin{document}\sloppy\sloppypar\raggedbottom\frenchspacing

\noindent
\title{A likelihood function for the \Gaia\ Data\footnote{%
  Copyright 2018 the author. Feel free to reproduce and redistribute, provided
  that you make no changes whatsoever.}}

\author[0000-0003-2866-9403]{David W. Hogg}
\affil{Center for Cosmology and Particle Physics, Department of Physics, New York University, 726~Broadway, New York, NY 10003, USA}
\affil{Center for Data Science, New York University, 60 Fifth Ave, New York, NY 10011, USA}
\affil{Max-Planck-Institut f\"ur Astronomie, K\"onigstuhl 17, D-69117 Heidelberg}
\affil{Flatiron Institute, 162 Fifth Ave, New York, NY 10010, USA}

\begin{abstract}\noindent
When we perform probabilistic inferences with the \Gaia\ Mission data,
we technically require
a \emph{likelihood function}, or a probability of the (raw-ish) data as a function
of stellar (astrometric and photometric) properties.
Unfortunately, we aren't (at present) given access to the \Gaia\ data
directly;
we are only given a Catalog of derived astrometric properties for the stars.
How do we perform probabilistic inferences in this context?
The answer---implicit in many publications---is that we should look at the
\Gaia\ Catalog as containing the \emph{parameters of a likelihood function}, or
a probability of the \Gaia\ data, conditioned on stellar properties,
evaluated at the location of the data.
Concretely, my recommendation is to assume
(for, say, the parallax) that the Catalog-reported
value and uncertainty are the mean and root-variance of a Gaussian
function that can stand in for the true likelihood function.
This is the implicit assumption in most \Gaia\ literature to date;
my only goal here is to make the assumption explicit.
Certain technical choices by the Mission team slightly invalidate
this assumption for \DRone\ (\TGAS), but not seriously. Generalizing beyond \Gaia,
it is important to downstream users of any Catalog products
that they deliver likelihood information about the fundamental data;
this is a challenge for the probabilistic catalogs of the future.
\end{abstract}

\keywords{%
astrometry
---
catalogs
---
methods: statistical
---
parallaxes
---\\
proper motions
---
stars: distances
}

\section*{~}\clearpage
\section{Introduction}
The age of \Gaia\ is also (perhaps coincidentally) the age of principled
probabilistic inference in astrophysics.
For this reason, the \Gaia\ (\citealt{gaia}) data are being used in many probabilistic
inferences (for example, \citealt{tri3, hawkins, sesar, lin}).
These probabilistic inferences take many forms, and have different levels
of hierarchical complexity, but all of them require that there be, at base,
a likelihood function, or a probability for the \Gaia\ data as a function
of model parameters.
One confusing question investigators face is: What constitutes the \Gaia\ data?
And how do I write a probability over it, when all I get to see is the
official Catalog release, with photometric and astrometric parameters and associated
uncertainties?

There is a standard answer, but in most inferences it appears only implicitly:
The inferences presume that the \Gaia\ Catalog entries can be used to construct
a likelihood function approximation, which is appropriate for use in inferences.
In almost all work so far, this likelihood function has been given a Gaussian form,
with mean and variance set to Catalog values (for example, this is clearly
stated in the introductory remarks in \citealt{tri2}).
The implicit assumptions underlying this choice are
what I am attempting to make explicit in this \documentname.

An investigator can take one of (at least) two attitudes towards the \Gaia\ data:
The investigator can think of the \Gaia\ Catalog as \emph{being the data}, in which case
the assumption is that the generative process for each \Gaia\ Catalog entry
itself is Gaussian.
Or the investigator can think of the \Gaia\ data as taking some raw form,
from which the Catalog has been derived (by, say, pipelines),
in which case the Catalog contains parameters of a Gaussian
\emph{approximation} to the likelihood function for those raw data.
It turns out it doesn't matter which attitude the investigator takes; the
proposal for the \Gaia\ likelihood function made here works in either case;
the two attitudes are identical in the limit that the Catalog contains
\emph{sufficient statistics} of the raw data.
Indeed, it is almost a definition of sufficient statistics that---if you have them---you
can use them to construct a good approximation to the likelihood function for the
raw data.
All that said, we are going to take the latter attitude: That is, that the
\Gaia\ data are raw data, and the Catalog is delivering statistics that can
be used to construct an approximation to the likelihood function.

One contemporary trend in astrophysics is to think about replacing rigid
catalogs with something more probabilistic, possibly representing uncertainties
through a sampling in catalog space (for example, \citealt{brewer, portillo}).
This idea is also informing some of the expected high-level outputs from
the \Gaia\ Mission too (\citealt{apsis}).
These ideas are interesting and new, and connect to the age of principled
probabilistic inference in which we find ourselves.
However, these ideas also come with substantial risk:
In many cases, a posterior sampling or posterior probability information
does not successfully encode sufficient likelihood information to permit
downstream analyses (with, say, different priors).
That is, investigators generally want---from an experiment or data source---likelihood
information, not posterior information.
This is because different investigators can have very different priors,
even qualitatively different priors; they won't agree on anything about the
data except what new information those data bring.
Also, if they want to combine $N$ stars in an inference, they obtain dangers
of taking some prior to the $N$th power if they can't get back to pure
likelihood information.
These concerns all flow from two principles: The first is the likelihood principle,
which states that new knowledge comes in likelihood form.
The second is the subjectivity of inference, or the principle that
\emph{an experiment ought to produce likelihood updates for all investigators,
no matter what their prior beliefs}.

\section{A likelihood function for \textsl{Gaia}}
In the simplest possible case, imagine that we are trying to infer
the true\footnote{There are many possible meanings for the word ``true''.
  In this context, we say the ``true distance'' because it is not the measured
  distance, but rather the distance that the star \emph{truly has} in some \emph{model}.}
distance $d_n$ to a star $n$ given the \Gaia\ data $y_n$ that
pertain to star $n$, and nothing else
(which is the goal, for example, of \citealt{tri2})
The inference looks like this:
\begin{eqnarray}
p(d_n\given y_n) = \frac{1}{Z_n}\,p(y_n\given d_n)\,p(d_n)
\label{eq:inference}
\quad ,
\end{eqnarray}
where
$p(d_n\given y_n)$ is the posterior pdf for the true distance $d_n$ given the data $y_n$,
$Z_n$ is a normalization constant,
$p(y_n\given d_n)$ is the likelihood (or the pdf for the data given the distance),
and $p(d_n)$ is the prior pdf for the true distance.
In order to perform this inference, we need the likelihood and the prior.

Note that, in order to write down these schematic equations,
we don't need to be perfectly specific here about what, exactly,
is the data $y_n$.
It could be all the raw \Gaia\ data pertaining to this particular star $n$,
or it could be the catalog entry in the \Gaia\ Catalog on star $n$.
Nothing about this formalism changes with these different choices,
although there might be some implicit marginalizations over nuisance
parameters in some of these particular cases.
That is, what you explicitly put in for the likelihood function $p(y_n\given d_n)$
will depend on what you consider to be the data $y_n$, but the formal structure
will be identical.

My goal here is to promote a particular choice for this likelihood function.
Getting straight to the point, in this simplest possible case, 
\begin{eqnarray}
p(y_n\given d_n) &=& p(y_n\given\varpi_n)
\label{eq:gotoparallax}
\\
\varpi_n &\equiv& \frac{1\,\AU}{d_n}
\\
p(y_n\given\varpi_n) &=& A_n\,N(\varpi_n\given\hat{\varpi}_n,\hat{\sigma}^2_{\varpi n})
\label{eq:onedlike}
\quad ,
\end{eqnarray}
where
$\varpi_n$ is the true parallax to star $n$ at true distance $d_n$ (implicitly $\varpi_n$ is measured in radians here),
$p(y_n\given\varpi_n)$ is the likelihood as a function of true parallax (rather than distance),
$A_n$ is a normalization (and units-conversion) constant,
$N(\xi\given\mu,V)$ is the Gaussian pdf for $\xi$ given mean $\mu$ and variance $V$,
$\hat{\varpi}_n$ is the (noisy) value given for the parallax of star $n$ in the \Gaia\ Catalog,
and $\hat{\sigma}_{\varpi n}$ is the value given for the uncertainty on that parallax.
The amplitude $A_n$ is not directly given in the \Gaia\ Catalog
but it turns out---because of that factor of $1/Z_n$ in
\equationname~(\ref{eq:inference})---it isn't needed for parameter-estimation-like
inferences.
If you \emph{do} find that you are doing an inference for which the amplitude
$A_n$ matters---for instance some inferences that might involve comparing certain
kinds of fully marginalized likelihoods---then $A_n$ can probably be reconstructed
from a goodness-of-fit statistic in the Catalog (that we expect in \Gaia\ \DRtwo).

It is more stable numerically to do inferences with logarithms of probabilities.
In the log,
\begin{eqnarray}
\ln p(y_n\given\varpi_n) &=& Q_n - \frac{1}{2}\,\frac{[\varpi_n - \hat{\varpi_n}]^2}{\hat{\sigma}^2_{\varpi n}}
\label{eq:onedlikelog}
\\
Q_n &\equiv& \ln\frac{A_n}{\sqrt{2\pi\,\hat{\sigma}^2_{\varpi n}}}
\quad .
\end{eqnarray}

This choice (\ref{eq:onedlike}) or (\ref{eq:onedlikelog}) for the likelihood function is Gaussian,
and presumes that the
Catalog values for the parallax and its uncertainty are accurate and represent
a likelihood maximum and width.
It is the choice made in many publications (for example, \citealt{tri2, leistedt, hawkins, lin}, among many others).
It has the property that the likelihood peaks when the true parallax
matches the Catalog-reported parallax,
and that the function is symmetric in parallax space---not distance space---because
\Gaia\ measures geometric parallaxes, not distances.
The likelihood function is a pdf for the data, evaluated at the data (which,
for a Bayesian, are fixed; \citealt{jaynes}), and therefore although it is a pdf over data,
it is really a function of the true parallax:
The likelihood function (\ref{eq:onedlike}) returns the answer to the question:
How probable are the observed data, if $\varpi_n$ is the true parallax of
star $n$?

It is worth making a few technical notes related to dimensions or units:
Because the data $y_n$ and
the true parallax $\varpi_n$ have (in general, though not always) different units,
the unknown amplitude $A_n$ will have (in general) non-trivial units.
Also, you might think
that the conversion from distance $d_n$ to parallax $\varpi_n$
in \equationname~(\ref{eq:gotoparallax})
would bring in some Jacobian factors of the form $||\dd d_n/\dd\varpi_n||$.
However, because the true distance only \emph{parameterizes} a function of the data,
or because the likelihood function has units of per-data (and not per-parallax),
the change of parameters doesn't bring a change of units, at least not in the
likelihood function (it would bring a change of units in the posterior pdf, or the
prior pdf).
I say more about these units issues elsewhere (\citealt{calculus}).

In a more general inference, it is not just the parallax (or distance)
that the investigator is modifying, but the (say) $D=5$ astrometric quantities
(celestial positions, proper motions, and parallax)
or a non-trivial subset of $D<5$ of these.
In this case the likelihood becomes
\begin{eqnarray}
p(y_n\given X_n) &=& A_n\,N(X_n\given\hat{X}_n,\hat{C}_{Xn})
\label{eq:like}
\quad ,
\end{eqnarray}
where now
$X_n$ is a $D$-vector of true values for the astrometric quantities for star $n$,
$p(y_n\given X_n)$ is the likelihood as a function of that vector of true quantities,
$A_n$ is again a normalization (and units-conversion) constant,
$N(\xi\given\mu,V)$ is now the Gaussian function for $D$-vector $\xi$ given $D$-vector mean $\mu$ and $D\times D$ covariance matrix $V$,
$\hat{X}_n$ is the $D$-vector of values given for the $D$ astrometric quantities for star $n$ in the \Gaia\ Catalog,
and $\hat{C}_{Xn}$ is the value given for the $D\times D$ covariance matrix for that
$D$-vector.
Again, the assumption here is that the likelihood function has Gaussian form, peaked
when the true values match the Catalog values, and symmetric in the quantities
reported in the Catalog (which are celestial positions, proper motions, and parallaxes).
In the log, this is
\begin{eqnarray}
\ln p(y_n\given X_n) &=& Q_n - \frac{1}{2}\,[X_n - \hat{X}_n]\T\cdot\hat{C}_{Xn}\inv\cdot [X_n - \hat{X}_n]
\\
Q_n &\equiv& \ln\frac{A_n}{\sqrt{||2\pi\,\hat{C}_{Xn}||}}
\quad ,
\end{eqnarray}
where we have implicitly assumed that the $D$-vectors are column vectors.

One comment to make briefly here is that although the \Gaia\ Catalog contains
the covariance matrices $\hat{C}_n$, those matrices are not in the data in a
trivial form.
They must be constructed from the Catalog entries.
Read the manual for details.

What about transformations of the astrometric quantities?
If the \Gaia\ likelihood is Gausssian in the Catalog values of celestial position,
parallax, and proper motion, is it also Gaussian in other functions of those
variables?
The answer is that it is only Gaussian in strictly linear transformations
of these parameters.
It isn't Gaussian, therefore, in distance (which is inverse parallax), nor
is it Gaussian in velocity (which is proper motion over parallax).
That is, the likelihood function in \equationname~(\ref{eq:onedlike}) has a
Gaussian form when plotted against parallax, but it doesn't when plotted against
distance.
That said,
The proper-motion likelihood will be Gaussian in any celestial coordinate
system you prefer (equatorial or ecliptic or Galactic, for example),
because these transformations (coordinate changes)
are simply rotations, which themselves are linear transformations.
Of course,
if you transform the $D$-vector $\hat{X}_n$ by a linear operator $R$ to make a new
vector $R\cdot\hat{X}_n$, you must also transform the covariance matrix to the new
matrix $R\cdot\hat{C}_{Xn}\cdot R\T$.

The likelihood functions of (\ref{eq:onedlike}) and (\ref{eq:like}) are
probabilities of the observed \Gaia\ data as a
function of astrometric parameters.
But there are many more parameters, including photometric, point-spread-function,
and spacecraft-attitude parameters as well, all of which contribute to the probability
for the data.
How do we deal with these nuisance parameters, or how can we ignore them?
Implicitly, I am assuming here that the \Gaia\ Catalog team has either optimized
or marginalized out these nuisance parameters.
Since the \Gaia\ Mission delivers immense signal-to-noise on these nuisance
parameters in most cases (\citealt{holl}), it won't matter much, technically, to this
discussion whether they optimize
the nuisance parameters or marginalize them out.

Of course, most inferences are not as simple as the inference described by
\equationname~(\ref{eq:inference}).
Except in rare cases, we aren't just trying to find out the distance to a single
star!
We usually are trying to fit some model of the Galaxy, or of some set of stars,
or calibrate a color-luminosity relationship, or something like that
(for example, \citealt{sesar, hawkins, leistedt, oh, delgado, widmark, anderson}, and many others).
In these cases, there is a model $p(X_n\given\theta)$ that says what we expect
for star $n$'s true astrometric quantites $X_n$, given a set of parameters $\theta$
of our larger model.
Within that larger model, the likelihood for a single star becomes
\begin{eqnarray}
p(y_n\given\theta) &=& \int p(y_n\given X_n)\,p(X_n\given\theta)\,\dd X_n
\label{eq:marginalize}
\quad ,
\end{eqnarray}
where we have marginalized out the true astrometric properties $X_n$ for star $n$.
These marginalized likelihoods $p(y_n\given\theta)$
can be multiplied together (or, better, added in the log)
to make likelihoods for collections of stars.
None of that hierarchical structure or marginalization changes the story here,
which is that the internal likelihood $p(y_n\given X_n)$ should be given the
Gaussian form in \equationname~(\ref{eq:like}).

Related to this, sometimes a model very specifically determines the
astrometric properties (celestial position, parallax, and proper
motions) of the star or a subset of those. Sometimes a model only
determines \emph{distributions} over those parameters.
If the model is deterministic in this sense, then the investigator can take
the $D$-slice (the subset of $D$) of the five astrometric parameters
in the \Gaia\ Catalog that are determined,
and the $D\times D$ sub-part of the $5\times 5$ covariance matrix
and use that in the likelihood form given in \equationname~(\ref{eq:like}).
An example of this is given by \citet{tri2}.
If the model is not deterministic, but rather produces distributions
over some set of parameters, then the investigator will have to do integrals
like that shown in \equationname~(\ref{eq:marginalize}).
An example of this is given by \citet{oh}.

Finally, you might want to do inference not purely on the astrometric
properties (celestial position, parallax, and proper motion) but also
on photometric properties (magnitudes, colors, reddenings).
In this \documentname, I don't take a position on the best likelihood function
approximations for these other observables, but in general my recommendations
would be similar: Use the Catalog entries to construct a Gaussian function
to stand in for the likelihood function.
In general, the magnitudes will have uncertainties that are closest to
Gaussian in flux space (not magnitude space) but I expect that the photometric
information from the Mission is so precise that it won't matter that much
what you assume here for most kinds of stars.

\section{Discussion}
In order to perform inferences
with the \Gaia\ data, we need a likelihood function.
The main point of this \documentname\ is that a sensible likelihood function
stand-in or surrogate can be constructed from the \Gaia\ Catalog, under the assumption
that the Catalog contains likelihood information, and accurate (and sufficient)
statistics of the data.
I give explicit forms for the likelihood function surrogate in
\equationname s~(\ref{eq:onedlike}) and (\ref{eq:like}).
These likelihood functional forms are not new---as I have emphasized,
they are used in multiple places
in the literature (for example, \citealt{tri2, leistedt, oh, delgado})---the point of
this is to make the
likelihood function assumptions explicit.

One of my soap-box issues is that inference is technically \emph{subjective}.
The investigator must make decisions about what constitutes the data,
and details of the model that generates those data, in order to make
inferences about the world.
In this case, I am recommending making certain decisions.
The first and most important is to decide that the likelihood function
has a Gaussian form, with mean and width set by \Gaia\ Catalog entries.
That is, assuming that the data are generated from the world by some mechanism
that is fundamentally Gaussian in the true parameters.
I am also implicitly making another recommendation, but it is a bit more
subtle:
I am recommending making the assumption that the \Gaia\ Catalog contains
nearly-sufficient statistics about the raw data, and that the likelihood function
constructed here is some kind of approximation for a probability for the raw data
from the Mission (which we currently don't get to see).
Another equally valid attitude to take is that the \Gaia\ Catalog
\emph{is}
the \Gaia\ data, and then this is just an assumption about how the world
(which is, apparently, full of true values for stellar positions and velocities;
oh and a \DPAC)
generated the \Gaia\ Catalog.
It is worthy of note and comment here that although I take the former view,
there is no change to anything we do if we take the latter view.
That is, this is a purely philosophical position.

Now, in detail, what have we really assumed about the \Gaia\ data?
That is, what assumptions would make the Gaussian form accurate?
There are a host of things to say here, so I will only say a few of the
most important things.
The first is that we are assuming that the \Gaia\ team is delivering
accurate results.
We are assuming that the Catalog is accurate, in terms of parameter
measurements and associated error variances and covariances.
Second, we have also assumed that the Catalog is a representation of likelihood
information.
That is, the \Gaia\ \DPAC\ has optimized a likelihood, or something very much
like that.
Thirdly, we have assumed that the likelihood is close to Gaussian in form.
The likelihood will be exactly Gaussian when the raw data are connected
to the model by Gaussian noise, and the model is linear.
These requirements, amazingly, \emph{are} close to being met in the \Gaia\ global solution.
But for central-limit-theorem-like reasons, any likelihood will become close
to Gaussian when the data are good enough.
So we are assuming something jointly about the simplicity of the model and
the data constraining it.

The second assumption---that the Catalog is a representation of likelihood
information---is in fact violated in detail by \Gaia\ \DRone\ \TGAS.
This catalog is a set of posterior inferences, one per star (\citealt{michalik, dr1}).
However, the posterior inference for the \Gaia\ data was made using a prior
built from the \Tycho\ likelihood outputs (with effectively very broad
or nearly flat priors),
so the \TGAS\ Catalog can be seen as providing (nearly) pure likelihood information,
but for the combined \Gaia +\Tycho\ data.
In principle, any user of the \TGAS\ Catalog should build a likelihood function
that is a ratio of the posterior Gaussian to the prior (or a difference
of logs)!
But this is a detail in the context of the very weakly informative priors
used in the construction of the \TGAS\ data set.
The weak priors mean that the posterior pdf shape is very similar to
(or perhaps identical to) the likelihood
function shape for these data (that is, the combined \Gaia +\Tycho\ data).
This gets more challenging in the future, because the \Gaia\ Collaboration will
probably have to use
more informative priors in the future for many Catalog entries.

Another assumption we have made is that we can treat each star indpendently.
That is, we can write down a likelihood for star $n$ without considering any
other star $n'$.
This assumption actually comprises two qualitatively different assumptions.
The first is that the inference of the \Gaia\ attitude model (and other nuisances)
induces no covariances between stars.
This is not true in detail, but becomes more true as the Mission proceeds
(\citealt{holl}).
The second is that the raw data can be separated cleanly between star $n$ and
star $n'$.
This isn't true for stars that overlap consistently on the focal plane; that is,
for stars that are closer than a few tenths of an arcsecond.
These close pairs (or K-tuples) of stars will remain covariant in the
Catalog, no matter how much data are taken.
For now we are ignoring all of these effects, because for most stars and projects
these are probably small effects, and the Catalogs themselves ignore
them anyway.

There are other sources of non-Gaussianity:
There are cosmic rays and there is binary (and triple)
contamination and stellar blending and crowding and so on; won't these
induce non-Gaussianities?
They will, probably, and there are empirical hints in the data
of such issues.
However, Gaussianity is the standard assumption across \Gaia\ science at
present (see, for example, all the previous citations for uses of a
\Gaia\ likelihood function), and it is the only sensible assumption in a
context in which we are given nothing more than a mean value and a variance
around that.

On that note: One thing I like to say about Gaussianity is that if you are given a mean
and a variance for a distribution, and you believe those statistics to be
accurate, then the Gaussian is the \emph{most conservative} assumption one
can make.
It is the maximum-entropy distribution with that mean and variance!
Of course this point is something of a red herring, because it is precisely
when the noise is non-Gaussian that it will be impossible to obtain an
accurate estimate of the mean and variance!
The point here is that the Gaussianity assumption in this work is strongly connected
to the assumption that the \Gaia\ Catalog is delivering \emph{accurate} statistics of
the data.

I have listed many assumptions here; it is worthy of note that most
assumptions can be \emph{tested}.
Testing assumptions about the likelihood function can be challenging,
but there are classes of objects, like quasars, and co-moving binary stars,
and Solar-System objects, that have very predictable astrometric properties,
or properties that are highly constrained by external physical considerations.
These objects can be used to test likelihood-function assumptions.

Right now, with the simple 5-parameter solutions for stars (that is, descriptions
in terms of just a celestial position, parallax, and proper motion), it is possible
to do likelihood optimization straightforwardly because the model is close to linear.
Once the space opens up to binary companions, exoplanets, and hierarchical systems,
there will be no straightforward way to optimize the likelihood, even for
a single \Gaia\ source, and the likelihood function will become
(in general) multi-modal and non-Gaussian.
The \Gaia\ \DPAC\ will have to face two issues:
One is how to explore the full space or distribution of parameters that is
consistent with the data.
The other is how to represent and deliver that information to the user
so that it is usable in inferences.

Even now, some of the post-processing of \Gaia\ \TGAS\ data to deliver
improved distance and parallax estimates are producing not likelihood outputs
but rather posterior outputs (for example, \citealt{tri3, leistedt, anderson}).
In order for these outputs to be usable in downstream inferences,
they have to be convertable back into likelihood (or marginalized-likelihood) form.
It isn't sufficient to say ``the prior of the next experiment will be the posterior
of the previous experiment'' because the down-stream user might have extremely
different goals, or highly informative external data. In either case, that user
needs likelihood information from the Mission.
When the \Gaia\ \DPAC\ starts doing binary-star astrometric orbit inferences, the likelihood
functions will become complex and multi-modal in the orbit-parameter space,
and they will be faced with challenging questions about how to represent this
information back to the user properly.
Many choices here are either very computationally expensive, very complex, or
limited in support (in the mathematical sense).

Finally, I want to end by answering the age-old question:
\emph{Why does the \Gaia\ Catalog contain negative parallaxes?}
The answer, in the context of these remarks, is that the Catalog
is a description of the likelihood function, and
for some stars, the peak of the Gaussian likelihood function happens
to be at a negative parallax!
In general, if a model is something like a linear fit in the presence of Gaussian
noise (and the \Gaia\ astrometric solution \emph{is} something like this), then
at low signal-to-noise the model will produce (at the maximum of its likelihood
function) negative linear coefficients just as readily as positive coefficients.
That is, linear fitting in a likelihood context will always produce negative
model parameters.
If the Catalog produced posterior information, rather than likelihood information,
these negative parallaxes might get cut off by a prior.
But for our purposes, this would be bad!
We want the Catalog to translate into a simple description
of the likelihood function so we can do simple inferences.
Those requirements put us in a world in which \Gaia\ reports many
negative parallaxes.

Thank you, \Gaia\ Mission, \Gaia\ \DPAC, and \Gaia\ Collaboration.

\acknowledgements
It is a pleasure to thank
  Lauren Anderson (Flatiron),
  Coryn Bailer-Jones (\acronym{MPIA}),
  Berry Holl (Geneva),
  Boris Leistedt (\acronym{NYU}),
  Lennart Lindegren (Lund),
  Hans-Walter Rix (\acronym{MPIA}),
and the attendees of the
  weekly New York City group meetings about Stars and about \Gaia\ \DRtwo\ preparation
for help with this project.
This work was partially supported by
  the National Science Foundation (grant \acronym{AST-1517237}).

This project implicitly (though not explicitly) made use of data from the
the European Space Agency
mission \Gaia\ (\url{http://www.cosmos.esa.int/gaia}), processed by the
\Gaia\ Data Processing and Analysis Consortium (\DPAC,
\url{http://www.cosmos.esa.int/web/gaia/dpac/consortium}). Funding for the
\DPAC\ has been provided by national institutions, in particular the
institutions participating in the Gaia Multilateral Agreement.

\bibliography{gaia}

\end{document}